\documentclass[amsmath,amssymb]{elsart}
\usepackage[latin1]{inputenc}
\usepackage{graphicx}
\usepackage{graphics}
\usepackage{amsfonts}
\begin{document}
\begin{frontmatter}

\title{On the quantum dynamics of a point particle in conical space}
  \author{C. Filgueiras and F. Moraes}
\address{Departamento de F\'\i sica, CCEN,  Universidade Federal 
da Para\'\i ba, Caixa Postal 5008, 58051-970 , Jo\~ao Pessoa, PB,
Brazil}
\begin{abstract}
A quantum neutral particle, constrained to move on a conical surface, is used as a toy model to explore bound states due to both a inverse squared distance potential and a $\delta$-function potential, which appear naturally in the model. These pathological potentials are treated with the self-adjoint extension method which yields the correct boundary condition (not necessarily a null wavefunction) at the origin. We show that the usual boundary condition requiring that the wavefunction vanishes at the origin is arbitrary and drastically reduces the number of bound states if used. The situation studied here is closely related to the problem of a dipole moving in conical space.
\end{abstract}
 \end{frontmatter}
%
  
\section{Introduction}
The simple, but nontrivial, geometry of the cone appears as an
 effective geometry in such diverse physical entities as cosmic strings
 \cite{shell}, defects in elastic media \cite{katanaev}, defects in
 liquid
 crystals \cite{caio} and so on. Accordingly, the dynamics of a quantum
 particle in a conical background has been profusely studied with very
 different motivations \cite{qcone}. An important issue concerning the cone is the fact that the conical background is naturally associated to a curvature 
singularity at the cone tip. The simplest way of dealing with this singularity is to impose 
the vanishing of the wavefunction at the cone tip as was done in \cite{qcone}. In fact, this is only one of the possible 
boundary conditions \cite{hagen}. A more general treatment can be done by use of the self-adjoint extension method \cite{symon}.
Apparently this leads to a family of boundary conditions but, in fact, only the boundary condition corresponding to the actual physics of the problem should hold. This was done, for example, in \cite{hagen,gerbert,alf}.

Another problem that requires self-adjoint extension involves a potential
 which
 goes with the inverse squared distance. This
 pathological potential has deserved some attention recently (see for
 example
 \cite{giri, patho} and references therein) although it has been
 addressed already in 1950 by K. M. Case \cite{case}. The main problem
 with
 this potential is that the energy levels are unbounded from below
 making
 the bound states unstable. Different schemes of regularization have
 been
 used to approach this problem including  radial cutoff \cite{carlos2}
 or self-adjoint extension \cite{giri}. In this article, we are interested in applying the self-adjoint extension 
approach to study the quantum dynamics of a neutral particle confined to a conical surface. As it will be seen below, besides 
the singularity at the cone tip, there is also a contribution from a inverse squared distance interaction.

When a quantum point particle moves confined to a surface embedded in  ordinary
 3-dimensional Euclidean space, it is subjected to a geometric potential \cite{rccosta}. 
It happens that, for the cone, part of this potential depends on the inverse squared distance from the 
cone tip. 

Using polar coordinates $\rho$ and $\theta$, we introduce the following
 line element 
\begin{equation}
ds^{2}=d\rho^{2}+\alpha^{2}\rho^{2}d\theta^{2}, 
\label{string1}
\end{equation}
such that $\rho\geq 0$ , $0 \leq \theta \leq 2\pi$. 
Metric (\ref{string1}) describes a cone if $0<\alpha<1$. Figure 1
 shows the making of a cone from a planar sheet where an angular
 section
 was removed with posterior identification of the edges. If $\gamma$ is
 the angle that defines the removed section then the remaining surface
 corresponds to an angular sector of $2\pi\alpha=2\pi-\gamma$. This is
 exactly what metric (\ref{string1}) describes. The incorporation of
 the
 term $\alpha^2$ to the planar metric in polar coordinates makes the
 total
 angle on the surface be $\int_{0}^{2\pi} \alpha d\theta=2\pi\alpha
 <2\pi$, since $0<\alpha<1$.  By identification of the length of the
 circle without the  sector, $2\pi\alpha\rho$, with the length of the
 complete
 circle it turns out to be on the cone, $2\pi\rho\tan\beta$, we get the
 relation
\begin{equation}
 \alpha=\tan\beta,
\end{equation}
where $2\beta$ is the opening angle of the cone (see figure 1). It is
 clear then that $\alpha$ tells how
 ``pointed'' is the cone. The closer $\alpha$ gets to 1 (or,
 equivalently, $2\beta$ to $\pi$) the flatter is the
 cone. For $\alpha=1$ the cone turns into a plane. If $\alpha>1$,
 relation (\ref{string1}) still holds and the conical surface
 corresponds to
 the insertion of a sector ({\it i.e.} $2\beta>\pi$). We call the
 resulting surface an anti-cone.
\begin{figure}[!h]
\begin{center}
\includegraphics[height=2cm]{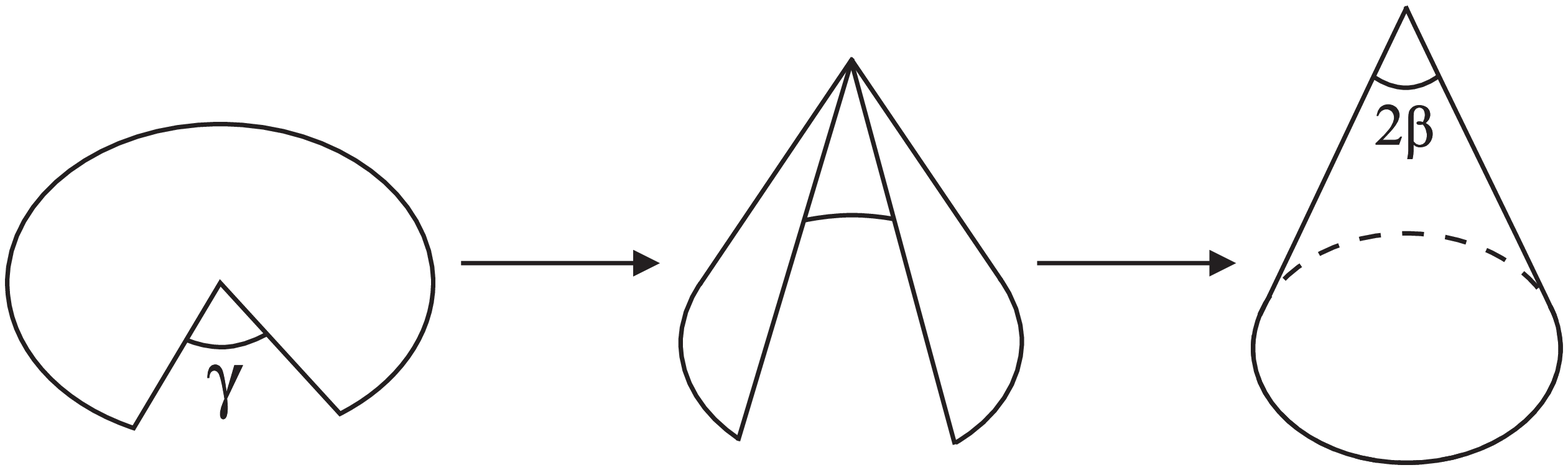} 
\caption{Conical surface of angular deficit $\gamma$.}
\end{center}
\end{figure}
  Notice that the line element (\ref{string1}) is just the $t=const.$,
 $z=const.$ section of the metric of the cosmic string spacetime
\begin{equation}
ds^{2}=c^2 dt^2-dz^2-d\rho^{2}-\alpha^{2}\rho^{2}d\theta^{2},
\label{cstring}
\end{equation}
where, in this case, $\alpha$ is related to the linear mass density
 $\mu$ of the string by $\alpha=1-4G\mu/c^2$, where $G$ is the
 gravitational constant and $c$ is the speed of light. Usually, only
 $\alpha<1$ is
 considered in cosmology, since $\alpha>1$ corresponds to a negative
 mass
 density string. For the general case we are treating here we consider
 both possibilities. We will see that this has important consequences
 on
 the number of bound states.

\section{The model}
Let us consider a neutral particle confined to a conical surface. As a consequence of the nontrivial topology of the cone and also because of two-dimensional confinement, the geometric potential should be taken into account \cite{rccosta}:
\begin{equation}
U_{geo}=-\frac{\hbar^{2}}{2M}\left(H^{2}-K\right),\label{geometric}
\end{equation}
where $H$ is the mean curvature and $K$ is the Gaussian curvature of
 the surface. For the cone \cite{caio2},
\begin{equation}
K=\left(\frac{1-\alpha}{\alpha}\right)\frac{\delta(\rho)}{\rho},
 \label{gauss1}
\end{equation}
and 
\begin{equation}
H=\frac{\sqrt{1-\alpha^{2}}}{2\alpha\rho}.
\end{equation}

It is clear that the $\delta$-function singularity in the Gaussian
 curvature corresponds to the tip of the cone which, from now on, we
 refer
 to as ``defect'' because of its topological defect characteristics.
 Also, depending on $\alpha$,  both curvatures 
will contribute with either  attractive or  repulsive potentials.

The neutral particle, with its motion confined to the conical surface, is therefore
subjected to a resultant potential given by
\begin{eqnarray}
 U_{res}=-\frac{\hbar^{2}}{8M}\left(\frac{1-\alpha^{2}}{\alpha^2\rho^2}\right)
+\frac{\hbar^{2}}{2M}\left(\frac{1-\alpha}{\alpha}\right)\frac{\delta(\rho)}{\rho}.
\end{eqnarray}

 Real systems have finite radius defects meaning that the curvature
 singularity is smoothed across the defect diameter. Therefore, let us
 consider a conical defect with a nucleus with radius $a$ which is small as
 compared
 to the overall dimension of the system. We replace the Gaussian curvature contribution
to the potential by a short-ranged potential supported inside the
 nucleus of the defect, that is, we have $U_{short}(\rho\geq a)=0$. 
The Schr\"odinger equation for the particle, in this case, is
\begin{eqnarray}
 -\frac{\hbar^{2}}{2M}\left[\frac{1}{\rho}\frac{\partial}{\partial
 \rho}\left(\rho \frac{\partial}{\partial \rho}\right)
+ \frac{1}{\alpha^{2}\rho^{2}}\frac{\partial^{2}}{\partial
 \theta^{2}}\right.
+\left.\left({\frac{1-\alpha^{2}}{4\alpha^2}}\right)\frac{1}{\rho^2}\right]\Psi+U_{short}(\rho)\Psi=E\Psi,\label{full}
\end{eqnarray}
where   $U_{short}(\rho)=\frac{\hbar^{2}}{2M}K$. Here, $K$ turns into (\ref{gauss1})
in the limit $a\rightarrow 0$.

  Following Kay and Studer \cite{kay}, we solve this problem by modelling it
 by boundary conditions. We  substitute the true problem above by 
\begin{eqnarray}
-\frac{\hbar^{2}}{2M}\left[\frac{1}{\rho}\frac{\partial}{\partial
 \rho}\left(\rho \frac{\partial}{\partial
 \rho}\right)+\frac{1}{\alpha^{2}\rho^{2}}\frac{\partial^{2}}{\partial
 \theta^{2}}\right.
+\left.\left({\frac{1-\alpha^{2}}{4\alpha^2}}\right)\frac{1}{\rho^2}\right]\Psi_{\eta}=E\Psi_{\eta},\label{ideal}
\end{eqnarray}
with $\Psi_{\eta}$ labeled by a parameter $\eta$ which is related to
 the behavior of the wavefunction in the limit $\rho\rightarrow a$. But
 we
 can not impose any boundary condition ({\it e.g.} $\Psi=0$ at $\rho=0$) without discovering which
 boundary conditions are allowed to equation (\ref{ideal}). This is the
 scope of the self-adjoint extension \cite{symon,boneau}.

We finish this section by remarking that the problem of a dipole in a conical background is qualitatively the same since the conical topology introduces a self-interaction which goes with the inverse squared distance to the cone vertex \cite{topological,carlos1,carlos2}. Moreover, the dependence of this self-interaction  on the cone opening angle is the same as the one that appears in the mean curvature contribution to the geometric potential. For these reasons, our results also shade some light onto the dipole problem \cite{giri}. 

\section{Self-adjoint extension}
  In order to proceed to the self-adjoint extension of (\ref{ideal}), we
  use the tensorial decomposition $L^{2}(R^{+},\rho d\rho)\otimes
 L^{2}(S^{1},d\theta)$. As we can see in \cite{point}, the operator
 $-\frac{\partial^{2}}{\partial \theta^{2}}$ is essentially
 self-adjoint in
 $L^{2}(S^{1},d\theta)$. Then, putting the wave function in the form
\begin{equation}
\Psi_{\eta}(\rho,\theta)=\Phi_{\eta}(\rho)e^{il\theta},\label{parati}
\end{equation} 
where $l=0,\pm 1,\pm 2$... is the angular momentum quantum number,
we arrive at the modified Bessel equation
\begin{equation}
\left[\frac{1}{\rho}\frac{d}{d
 \rho}\left(\rho \frac{d}{d
 \rho}\right)-\left(\frac{\nu^{2}}{\rho^{2}}+k^2\right)\right]\Phi_{\eta}=0,
 \label{dipocone}
\end{equation}
where $k^2=-\frac{2ME}{\hbar^2}>0$, since we are looking for bound states, and
with
\begin{equation}
\nu^{2}=\frac{l^{2}}{\alpha^{2}}-\frac{\left(1-\alpha^{2}\right)}{4\alpha^{2}}.
 \label{nu2}
\end{equation}
Notice that  $\alpha>1$ implies $\nu^2>0$ for all allowed values of $l$. On the other hand, when
 $\alpha<1$ we have $\nu^2<0$ for $l=0$ and $\nu^2>0$ for $l=\pm 1,\pm 2$... There is no choice of $l$ that will give $\nu=0$,
 except $l=0$, but then $\alpha$ would have to be $1$ (flat space). 
  
Now, to find the full domain of $\Phi_{\eta}$ in $L^{2}(R^{+},\rho
 d\rho)$, we have to find the deficient subspace of (\ref{dipocone}).
 To do
 this, we have to solve the eigenvalue equation
\begin{equation}
\mathcal{H}^{\dagger}\Phi_{\pm}=\pm ik_{0}\Phi_{\pm}, \label{eigen}
\end{equation}
where $\mathcal{H}=\left[\frac{1}{\rho}\frac{d}{d
 \rho}\left(\rho \frac{d}{d
 \rho}\right)-\left(\frac{\nu^{2}}{\rho^{2}}+k^2\right)\right]$ comes from equation (\ref{dipocone}),
for each case: $\nu^2<0$ and $\nu^2>0$.

The only square integrable functions which are solutions to  equation
 (\ref{eigen}) are the modified Bessel functions $K_{\mu}$ such that
\begin{equation}
\Phi_{\pm}(\rho)=const.
 K_{\mu}\left(\frac{\rho}{\hbar}\sqrt{\mp2iMk_{0}}\right),
 \label{subspace}
\end{equation}
where $\mu=\nu$ if $\nu^2>0$ or $\mu=i|\nu|$ if $\nu^2<0$.
The dimension of such deficient
 space is $(n_{+},n{-})=(1,1)$. Because of this, the domain of
 (\ref{dipocone}) in $L^{2}(R^{+},\rho d\rho)$ is given by the set of
 functions
\begin{eqnarray}
\Phi_{\eta}(\rho)=\chi_{\mu}(\rho)+C\left[K_{\mu}\left(\frac{\rho}{\hbar}\sqrt{-2iMk_{0}}\right)\right.
+\left.e^{i\eta}K_{\mu}\left(\frac{\rho}{\hbar}\sqrt{2iMk_{0}}\right)\right],\label{domain}
\end{eqnarray}
where $\chi_{\mu}(\rho)$, with
 $\chi_{\mu}(a)=\dot{\chi}_{\mu}(a)=0$, is the wavefunction when we do
 not have
 $U_{short}(\rho)$. The last term in (\ref{domain})  gives the correct behavior of the
 wavefunction when $\rho=a$. The parameters $\eta(mod2\pi)$ and $k_0$ represent the {\it a priori} choices of boundary conditions. As we shall see below, the physics of the problem determines these parameters without ambiguity. In fact, $k_0$ cancels out of the calculations such that we only have to determine $\eta$, which
 describes the coupling between $U_{short}(\rho)$ and the wavefunction.
 Then
 it must be expressed in terms of  $\alpha$, the defect
 core radius $a$ and the effective angular momentum $\nu$. The next
 step
 is to find a fitting to $\eta$ compatible with $U_{short}(\rho)$. We
 shall do
 this in the next two sections for the $\nu^{2}<0$ and $\nu^{2}>0$
 cases, respectively.
\section{Case $\nu^{2}<0$}
There is only one possibility here, which is $l=0$ and $\alpha<1$. Now,
\begin{equation}
U_{eff}=\frac{\hbar^{2}}{2M}\frac{l^2}{\alpha^2\rho^2}-\frac{\hbar^{2}}{8M}\left(\frac{1-\alpha^{2}}{\alpha^2\rho^2}\right)
\end{equation}
is the effective potential that includes the centripetal term and the contribution from the mean curvature. The contribution from the Gaussian curvature to the potential is
\begin{equation}
U_{short}=\frac{\hbar^{2}}{2M}\left(\frac{1-\alpha}{\alpha}\right)\frac{\delta(\rho)}{\rho}.
\end{equation}
Then, in this case, 
$U_{eff}<0$ and therefore attractive; $U_{short}>0$, repulsive. We will see below that, even with this very short range repulsion, the attractive $1/\rho^2$ potential guarantees a bound state.

In this section we  find a fitting formula for $\eta$ following the
 procedure described by Kay and Studer \cite{kay}. First, we write
 (\ref{full}), the true problem, for $\Phi_{static}^{true}(\rho)$, the
 $E=0$ or
 static solution: 
\begin{equation}
\left\{-\frac{\hbar^{2}}{2M}\left[\frac{1}{\rho}\frac{\partial}{\partial
 \rho}\left(\rho
\frac{\partial}{\partial\rho}\right)-\frac{\nu^{2}}{\rho^{2}}\right]+U_{short}(\rho)\right\}\Phi_{static}^{true}=0
 \label{statictrue}
\end{equation}
  
Since we are considering the defect core radius $a$, the Gaussian
 curvature (\ref{gauss1}) must be written as
  \begin{equation}
K=\lambda\left(\frac{1-\alpha}{\alpha}\right)\frac{\delta
 \left(\rho-a\right)}{a}, \label{gauss2}
\end{equation}
where the constant $\lambda$ was inserted for convenience - for the
 case we are studying it is in fact 1, but if we make
 $\lambda\rightarrow\infty$ we are choosing the boundary condition $\Psi(\rho=a)=0$.
  Then, we require that 
\begin{equation}
\frac{\rho}{\Phi_{static}^{true}}\frac{d\Phi_{static}^{true}}{d\rho}|_{\rho=a}=\frac{\rho}{\Phi_{\eta,static}}\frac{d\Phi_{\eta,static}}{d\rho}|_{\rho=a},\label{fit}
\end{equation}
where $\Phi_{\eta,static}(\rho)$ comes from (\ref{domain}). For
 $\nu^{2}>0$ the wavefunctions (\ref{domain}) are given in terms of $K_{i|\nu|}$, the modified
 Bessel function  \cite{abram}  of purely imaginary order.
 Since $a\approx 0$ we use the expansion  for small $x$,
\begin{equation}
K_{i|\nu|}(x)\approx \sqrt{\frac{\pi}{\nu \sinh(\pi\nu)}}\sin\left[\nu
 \ln(x/2)+\nu\gamma\right]\left[1+O(x^2)\right]\label{nu},
\end{equation}
where $\gamma$ is the Euler-Mascheroni constant. Now, taking into
 account (\ref{domain}), we arrive at
\begin{eqnarray}
a\frac{d\Phi_{\eta,static}/d\rho}{\Phi_{\eta,static}(\rho)}|_{\rho\rightarrow
 a}=a\frac{\dot{F}_{\eta}(\rho=a)}{F_{\eta}(\rho=a)}\label{derivadaleft}
\end{eqnarray}
where 
\begin{eqnarray}
F_{\eta}(\rho)=\sin\left[\nu
 \ln\left(\sqrt{-2Mik_{0}}\rho/2\hbar\right)+\nu\gamma\right]
+e^{i\eta}\sin\left[\nu
 \ln\left(\sqrt{+2Mik_{0}}\rho/2\hbar\right)+\nu\gamma\right] \label{F}
\end{eqnarray}
and $\dot{F}_{\eta}=\frac{dF_{\eta}}{d\rho}$.
Now, integrating (\ref{statictrue}) from $0$ to $a$ we have
  \begin{eqnarray}
 a\frac{d\Phi_{static}^{true}\left(\rho=a\right)}{d\rho}= 
\int^{a}_{0}\rho
 d\rho\lambda\left(\frac{1-\alpha}{\alpha}\right)\frac{\delta
\left(\rho-a\right)}{a}\Phi_{true}^{static}(\rho) 
-\int_{0}^{a}\frac{\nu^{2}}{\rho^{2}}\Phi_{static}^{true}\rho d\rho.
\end{eqnarray}
Considering that 
\begin{eqnarray}
\int_{0}^{a}\frac{\nu^{2}}{\rho^{2}}\Phi_{static}^{true}(\rho)\rho
 d\rho\approx\frac{\nu^{2}}{a^{2}}\Phi_{static}^{true}(\rho=a)\int_{0}^{a}\rho
 d\rho,\nonumber
\end{eqnarray}
we have
\begin{equation}
\frac{a}{\Phi_{static}^{true}(\rho=a)}\frac{d\Phi_{static}^{true}}{d\rho}|_{\rho=a}=\lambda\left(\frac{1-\alpha}{\alpha}\right)-\frac{\nu^{2}}{2}\label{derivadaright}.
\end{equation}
  So, from (\ref{fit}), (\ref{derivadaleft}) and (\ref{derivadaright}),
 we obtain the relation
\begin{eqnarray}
a\frac{\dot{F}_{\eta}(\rho=a)}{F_{\eta}(\rho=a)}\approx\lambda\left(\frac{1-\alpha}{\alpha}\right)-\frac{\nu^{2}}{2},\label{firstfit}
\end{eqnarray}
which gives us the parameter $\eta$ in terms of the physics of the
 problem, that is, the correct behavior of the wave functions when
 $\rho\rightarrow a$, or the coupling between the short-ranged potential
 $U_{short}(r)$ and the wavefunctions. Next, we will find the bound states of the
 Hamiltonian and we will see that the formula (\ref{firstfit}) gives us
 the spectrum without any arbitrary parameter. For that, we must solve the eigenvalue problem  
\begin{equation}
-\frac{\hbar^{2}}{2M}\left[\frac{1}{\rho}\frac{\partial}{\partial
 \rho}\left(\rho \frac{\partial}{\partial
 \rho}\right)+\frac{\nu^{2}}{\rho^{2}}\right]\Phi_{E}=-E\Phi_{E}, \label{eigenvalue}
\end{equation}
whose general solution is given by
\begin{equation}
\Phi_{E}(\rho)=K_{i|\nu|}\left(\frac{\rho}{\hbar}\sqrt{-2mE}\right).
\end{equation}
Since this solution belongs to the domain of the Hamiltonian that appears in equation (\ref{dipocone}), it is of the form (\ref{domain}), that is,
\begin{eqnarray}
\Phi_{E}(\rho)=\chi_{\nu}(\rho)+C\left[K_{i|\nu|}\left(\frac{\rho}{\hbar}\sqrt{-2iME}\right)\right.
+\left.e^{i\eta}K_{i|\nu|}\left(\frac{\rho}{\hbar}\sqrt{2iME}\right)\right].\label{domain2}
\end{eqnarray}
 Using the expressions (\ref{domain2}) and (\ref{nu}), we arrive at
\begin{eqnarray}
F_{\eta}(\rho=a)= \sin\left[\nu\ln\left(\sqrt{-2ME}a/2\hbar\right)+\nu\gamma\right]
\end{eqnarray}
and
\begin{eqnarray}
\dot{F}_{\eta}(\rho=a)=\frac{\nu}{a} \cos\left[\nu\ln\left(\sqrt{-2ME}a/2\hbar\right)+\nu\gamma\right].
\end{eqnarray}

Using the above expressions for $F_{\eta}$ and $\dot{F}_{\eta}$ in (\ref{firstfit}) we get
  \begin{eqnarray}
\nu \cot\left[\nu
 \ln\left(\sqrt{-2ME}a/2\hbar\right)+\nu\gamma\right]=\lambda\left(\frac{1-\alpha}{\alpha}\right)-\frac{\nu^{2}}{2}.
\label{cot}
\end{eqnarray}
With $\lambda=1$, the inversion of equation (\ref{cot}) yields
\begin{equation}
E=-\frac{2\hbar^{2}}{Ma^{2}}exp\left[\frac{2}{\nu}\cot^{-1}\left(\frac{1-\alpha}{\alpha\nu}-\frac{\nu}{2}\right)-2\gamma\right].\label{novel}
\end{equation}

Notice that the case studied in this section,
 $\nu^2<0$, corresponds to $\alpha<1$. This means that, while the  mean curvature contributes attractively,  the Gaussian  curvature  contributes with a repulsive short-ranged
 potential.  Equation (\ref{nu2}) implies that the only allowable value for the angular momentum is $l=0$, meaning that we have a single bound state.

If we make
 $\lambda\rightarrow\infty$ in equation (\ref{cot})  we obtain
 the result of reference \cite{carlos2}:
\begin{equation}
E=-\frac{2\hbar^{2}}{Ma^{2}}\exp\left[-\frac{2n\pi}{\nu}-2\gamma\right]\label{moraes},
\end{equation}
with $n=1,2,...$, which corresponds to imposing the condition that the
 wavefunction vanishes at $\rho=a$, or an infinitely high barrier there.
  
\section{Case $\nu^{2}>0$}
Here, we have two possibilities: 
\begin{itemize}
 \item $\alpha<1$ and $l\neq0$ $\Longrightarrow$ $U_{eff}>0$, $U_{short}>0$ 
\item $\alpha>1$ and $\mbox{any}\ l$ $\Longrightarrow$ $U_{eff}>0$, $U_{short}<0$
\end{itemize}
In the first case, even though the contribution from the mean curvature to $U_{eff}$ is attractive,  $U_{eff}$ itself is not, then there are no bound states. In the second case, the attractive $\delta$-function potential guarantees one bound state for specific values of $l$, as it will be seen below. 

In this case  equation (\ref{statictrue}) is written as 
  \begin{equation}
\left\{-\frac{\hbar^{2}}{2M}\left[\frac{1}{\rho}\frac{\partial}{\partial
 \rho}\left(\rho \frac{\partial}{\partial
 \rho}\right)-\frac{\nu^{2}}{\rho^{2}}\right]+U_{short}(\rho)\right\}\Phi_{static}^{true}=0.
 \label{xinegative}
\end{equation}
  Then, solving  equation (\ref{eigen}), we arrive at the set of
 functions 
\begin{eqnarray}
\Phi_{\eta}(\rho)=\chi_{\nu}(\rho)+C\left[K_{\nu}\left(\frac{\rho}{\hbar}\sqrt{-2iMk_{0}}\right)\right.
+\left.e^{i\eta}K_{\nu}\left(\frac{\rho}{\hbar}\sqrt{2iMk_{0}}\right)\right].\label{domainxi}
\end{eqnarray}
These functions are square integrable only in the range  $\nu\in(-1,1)$ but, since we can not have $\nu=0$, we are restricted to $0<\nu^2<1$.
  
Now, following the procedure for the case $\nu^{2}<0$ above, and taking
 into account that \cite{abram}, for $\nu\neq0$,
\begin{eqnarray}
K_{\nu}(x)\rightarrow
 \frac{\pi}{2sin\left(\pi\nu\right)}\left[\frac{1}{\Gamma(-\left|\nu\right|+1)}\left(\frac{x}{2}\right)^{-\left|\nu\right|}\right.
 \left.
 +\frac{1}{\Gamma(\left|\nu\right|+1)}\left(\frac{x}{2}\right)^{\left|\nu\right|}\right]
\end{eqnarray}
when $\left|x\right|\rightarrow 0$, we find the energy spectrum to be
\begin{eqnarray}
E=-\frac{2\hbar^{2}}{Ma^{2}}\left[\frac{\Gamma(\left|\nu\right|+1)}{\Gamma(-\left|\nu\right|+1)}\left(\frac{1+\frac{1-\alpha}{\alpha\left|\nu\right|}+\frac{\left|\nu\right|}{2}}{1-\frac{1-\alpha}{\alpha\left|\nu\right|}-\frac{\left|\nu\right|}{2}}\right)\right]^{1/\left|\nu\right|}
\end{eqnarray}
where the dependence of the bound states on the angular momentum $l$ is analyzed below.

In the case studied in this section we have two constraints: $0<\nu^2<1$ and $\alpha>1$. The first one, combined with equation (\ref{nu2}), implies that
\begin{equation}
 \frac{1-\alpha^2}{4}<l^2<\frac{3\alpha^2+1}{4}.
\end{equation}
The inequality on the left hand side is always obeyed since $\alpha>1$. The right hand side inequality gives us the maximum value of $l$ for each value of $\alpha$. In other words, we  have bound states for\\
\begin{eqnarray}
 &1&<\alpha<\sqrt{5} \,\, \mbox{at}\,\, l=0 \,\,\mbox{and}\,\, \pm1,\nonumber\\
 &\sqrt{5}&<\alpha<\sqrt{35/3}\,\, \mbox{at}\,\, l=0, \pm1 \,\,\mbox{and}\,\, \pm2,\nonumber\\
 &\sqrt{35}&<\alpha<\sqrt{63/3} \,\,\mbox{at}\,\, l=0, \pm1, \pm2 \,\,\mbox{and}\,\, \pm 3,
\end{eqnarray}
 and so on.

Notice that the case studied in this section, $\nu^2>0$, corresponds to the quantum problem in the anti-cone ($\alpha>1$). This makes the  mean curvature potential repulsive and the short-ranged potential, due to the Gaussian curvature, attractive. Therefore, the Gaussian curvature is the sole responsible for the bound states. Notice also that, the arbitrary  boundary condition, $\Psi=0$ at $\rho=0$, used for example in references \cite{carlos1,carlos2,khali}, is not allowed since $K_{\nu}$ is not regular there. In other words, if this boundary condition is used there are no bound states for $\alpha>1$.

\section{Concluding remarks}

Many authors have discussed separately the quantum dynamics of a particle in the presence of spacial singularities or else, ill-behaved potentials, like the inverse squared distance potential and the $\delta$-function potential. In a single toy model we study these anomalous conditions in a unified way which clarifies the meaning of the boundary conditions usually taken for granted in such problems, like imposing that the wavefunction vanishes at the origin, for example. In fact, this imposition is quite arbitrary. The self-adjoint extension introduces a natural way of finding the appropriate boundary condition which describes the physics of short-ranged potentials. A common interpretation \cite{article} of the self-adjoint extension is that it gives a family of solutions associated to a certain freedom on choosing the boundary conditions. However, by considering a finite-sized defect and shrinking its radius to zero we fixed the boundary condition,  as done in \cite{alf}.

Our toy model consists in a neutral particle in a two-dimensional conical surface.  The conical geometry  introduces an inverse squared distance potential due to the mean curvature which can be either attractive or repulsive, depending on the cone parameter $\alpha$. The conical geometry is also responsible for a $\delta$-function interaction which, again, can either be attractive or repulsive, depending on the cone parameter $\alpha$.

Our results are summarized in Table I below.
  
\begin{center}
\begin{tabular}{|c|c|c|}
\hline
 &\multicolumn{1}{c|}{$\alpha>1$}&\multicolumn{1}{c|}{$\alpha<1$}\\
\hline
$\nu^{2}<0$ & $-----$ & 1 bound state for $l=0$\\ 
\hline
$\nu^{2}>0$ &  bound states & scattering states for $l\neq0$\\
\hline
\end{tabular}
\newline
\newline
Table I. Summary of the results. 
\end{center}

Table I reveals that, even in the case of a repulsive effective potential ($\alpha>1$), the attractive short-ranged potential guarantees  bound states for the values of the angular momentum specified in (40). Conversely, when the short-ranged potential is repulsive ($\alpha<1$)  an attractive effective potential potential assures one bound state ($l=0$). And, when $\alpha<1$, for $l\neq 0$, we have both $U_{eff}$ and $U_{short}$ repulsive, giving no bound states.

  Since our Schr\"odinger equation is singular at $\rho=a$ we 
 solved the 
problem paying attention to the correct behavior of the wavefunction
 there.
  Using the self-adjoint extension of the Hamiltonian operator this
 behavior appears naturally, but we had to fit the extension parameter in terms of the known physics of the system when
 $\rho\leq a$. 
 This procedure gave us exact analytical
 expressions for the energy levels. Furthermore, we showed that the ``usual'' boundary condition $\Psi=0$ at $\rho=a$, which corresponds to an infinite barrier there, gives bound states only for the case $\alpha>1$, or the anti-cone.

Without using the self-adjoint
 extension
 we can follow the usual way to deal with delta-potentials (see for example \cite{hagen}). However,
 the self-adjoint extension approach is a
 more direct procedure to find the physics compatible with short-ranged
 potentials. The reader can compare, for example, the
 Aharonov-Bohm-Coulomb problem discussed by Hagen and                Park \cite{hagen}
 and  by Park and Oh \cite{park}. The same non-relativistic spectrum was achieved in both articles but with much 
less work in the second, where the self-adjoint extension was used. Another example of the importance
of the method is given in reference \cite{article} where we studied the
gravitational bound-state Aharonov-Bohm effect due to a cosmic string. This effect was initially \cite{valdir} 
predicted for the hypothetical case of a cosmic string surrounded by a cylindrical wall. By using the correct 
boundary condition, as given by the self-adjoint extension, we  were able to show in \cite{article} that the effect is still there 
without the need of the wall. In summary, the self-adjoint extension 
 approach is a powerfull method if we are interested in the quantum
 dynamics  of particles in spaces with singularities.

{\bf Acknowledgments}\\ 

This work was partially supported by PRONEX/FAPESQ-PB,
 CNPq and CAPES (PROCAD). We are indebted to Prof. Mark Alford for his comments on reference \cite{article} and for pointing out to us reference \cite{alf}.

\end{document}